\documentclass[a4paper, aps, prd, final, reprint, showkeys,showpacs, superscriptaddress, floatfix, nofootinbib, twocolumn, amsmath, amssymb]{revtex4-1}
\usepackage[english]{babel}
\usepackage[utf8]{inputenc}
\usepackage[colorinlistoftodos, color=green!40, prependcaption]{todonotes}
\usepackage{multirow}

\usepackage{amsthm}
\usepackage{mathtools}
\usepackage{physics}
\usepackage{xcolor}
\usepackage{graphicx}
\usepackage[left=23mm,right=13mm,top=35mm,columnsep=15pt]{geometry} 
\usepackage{adjustbox}
\usepackage{placeins}
\usepackage[T1]{fontenc}
\usepackage{lipsum}
\usepackage{csquotes}
\usepackage{textgreek}
\usepackage{subfig}
\usepackage{paralist}
\usepackage{siunitx}
\usepackage[pdftex, pdftitle={Article}, pdfauthor={Author}]{hyperref} 
\usepackage{lineno}
\begin{document}
\title{Ionisation quenching factors from W-values in pure gases for rare event searches }

\author{I. Katsioulas}
\email[Correspondence email address: ]{i.katsioulas@bham.ac.uk}
\affiliation{School of Physics and Astronomy, University of Birmingham, B15 2TT, United Kingdom}

\author{P. Knights}
\affiliation{School of Physics and Astronomy, University of Birmingham, B15 2TT, United Kingdom}
\affiliation{IRFU, CEA, Universite Paris-Saclay, F-91191 Gif-sur-Yvette, France}

\author{K. Nikolopoulos}
\affiliation{School of Physics and Astronomy, University of Birmingham, B15 2TT, United Kingdom}


\begin{abstract}
  The effect of ionisation quenching for ions is critical for
  experiments relying on the measurement of low energy recoils, such
  as direct Dark Matter searches.  We present ionisation
  quenching factor estimates over a range of energies for protons, $\alpha$-particles, and heavier
  ions in H$_{2}$, CH$_{4}$, N$_{2}$, Ar, CO$_{2}$, and C$_{3}$H$_{8}$
  gases, estimated from the respective reference W-value
  measurements. The resulting ionisation quenching factors are compared
  with predictions from SRIM.
\end{abstract}

\maketitle

\section{Introduction} 
\label{sec:outline}

Direct searches for Dark Matter (DM) particles employ a range of detector technologies, including liquid noble gases~\cite{Aprile:2018dbl, Akerib:2016vxi, lux2016, zeplin2012}, crystals~\cite{The_CRESST_Collaboration2015-pz}, and semiconductors~\cite{supercdms2014}.  %
Gaseous detectors are used to expand DM searches below the Lee-Weinberg mass bound of about $2\;{\rm GeV}$~\cite{Lee:1977ua}, through the use of light-nuclei gases,
and for directional DM searches~\cite{Iguaz_2011, DAW2012397}. 

The NEWS-G experiment,  using a spherical proportional counter filled with a neon-methane gas mixture~\cite{Arnaud2018-nr}, set the most stringent exclusion limit at the time in the DM-nucleon spin-independent interaction cross section for a $500\;\si{\mega\eV}$ mass DM candidate. 
The TREX-DM experiment~\cite{Castel:2018gcp}, using a gaseous Time Projection Chamber (TPC) with argon-isobutane and neon-isobutane gas mixtures, is projected to achieve competitive sensitivities in the DM candidate mass region below a few $\si{\giga\eV}$.     
Directional DM searches such as DRIFT~\cite{Battat:2016xxe}, CYGNUS~\cite{Vahsen:2020pzb}, MIMAC~\cite{Iguaz_2011} and NEWAGE~\cite{Ikeda:2021ckk} employ low-pressure (below a few hundred millibar) TPCs with the ability to infer the ionisation track direction. Additionally, many of these detectors utilise gas mixtures containing odd-proton, e.g. $^{19}$F, or odd-neutron isotopes, e.g. $^{3}$He, to achieve world-leading sensitivity to spin-dependent DM-nucleon interactions.  

Knowledge of the expected amount of ionisation and scintillation induced by nuclear recoils is important for these searches, which rely on the detection and measurement of nuclear recoils with energy from a few eV to hundreds of keV, induced by elastic scattering of DM particles on nuclei. As a result, several investigations of the corresponding ionisation quenching effects have been performed~\cite{Chavarria:2016xsi,Scholz:2016qos,Sarkis:2020soy, Joo:2018hom, xenon_qf, silicon_qf, germanium_qf}.

Low energy nuclear recoils dissipate energy in a medium through inelastic Coulomb interactions with atomic electrons, referred to as electronic energy losses, 
and elastic scattering in the screened electric field of the nuclei, referred to as nuclear energy losses. Electronic energy losses dominate for fast ions, while nuclear energy losses are increasingly important for decreasing ion kinetic energy $E_{R}$, and become dominant
for ion velocities smaller than the electron orbital
velocity~\cite{evans2003atomic}. 
The produced secondary recoil atoms and electrons may also undergo scattering, further transferring energy to other particles.
 
The ionisation quenching factor, $q_f$, concept was 
introduced by Lindhard et~al.~\cite{lindhard1963integral}, and remains a field of intense study~\cite{Sorensen:2014sla}. The
electronic and nuclear stopping power are estimated 
assuming they are uncorrelated. Integrating the electronic and nuclear energy losses until the ion stops provides the energies $v$ and $\eta$ given to atomic motion and electrons, respectively. The maximum available energy for ionisation and scintillation is $\eta$. The Lindhard quenching factor is defined as $\eta/E_R$, the ratio of energy given to electrons over the ion kinetic energy.

\begin{figure}[!h]
\centering
\includegraphics[width=0.91\linewidth]{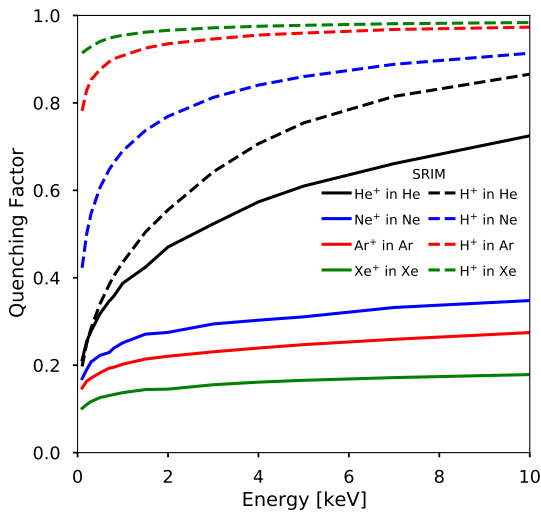}
\caption{Quenching factor for ions in their own gas
  estimated using SRIM~\cite{ziegler2008srim}.\label{fig:qf_srim}}
\end{figure}

SRIM~\cite{ziegler2008srim} (Stopping and Range of Ions in Matter) is a computational tool primarily used for the estimation of ion ranges in matter. TRIM (TRansport of Ions in Matter), a module of SRIM, is often used to estimate the ionisation quenching factor in  materials as it provides estimates for the fraction of ion kinetic energy dissipated in electronic and nuclear energy losses, including contributions of secondary recoils. 
SRIM estimates of the ionisation quenching factor of heavy ions in their own gas are presented in Fig.~\ref{fig:qf_srim}.
%

%

%

In the literature, multiple definitions of the ionisation quenching factor may be found depending on the application: 
\begin{enumerate}
\item the fraction of the ion kinetic energy that is
   dissipated in a medium in the form of ionisation electrons and
   excitation of atomic and quasi-molecular states;
\item the ratio of the ``visible'' energy in an ionisation detector to
   the recoil kinetic energy; and
\item the conversion factor between the kinetic energy of an electron and that of an ion that result to the same ``visible'' energy in the ionisation detector.
\end{enumerate}

In the above, the first definition coincides with that of Lindhard et~al. Moreover, the term ``visible'' refers to energy dissipated in a detectable form for the specific detection system, i.e. ionisation for detectors measuring charge.

For the experimental determination of the ionisation quenching factor in a material, the amount of ionisation induced by an ion species has to be measured as a function of its kinetic energy and compared to the corresponding ionisation by electrons of the same initial energy. In this case the ionisation quenching factor takes the form, 
\begin{equation}
    q_{f} = E_{d}/E_{R}\;,
    \label{eq:definitionBQuenchingFactor}
\end{equation}
where $E_{d}$ is the amount of electronic energy losses ``visible'' to
a detector~\cite{Iguaz_2011, DAW2012397}. 

For these measurements, ions with precisely known kinetic energy are required, along with the capability to measure small energy deposits. As a result, although the ionisation quenching factor is critical for the modelling of the detector response to nuclear recoils --- and, thus, for the sensitivity to discover DM --- relevant measurements are scarce for low energy ions in gases.
Santos et~al.~\cite{santos2008ionization}, investigated the quenching of He$^+$ in He/C$_4$H$_{10}$ gas mixtures using a table-top low energy ion-electron accelerator~\cite{Muraz:2016upt}.
The measurements were performed with a  
Micromegas proportional counter~\cite{GIOMATARIS2006405}. %
The ionisation quenching factor was estimated as the ratio
of the energy measured in the detector for ions to their initial total kinetic
energy. Beyond measurements, prompted by the requirements of the DM community, Hitachi estimated the dependence of the quenching factor on energy in liquid noble gas detectors~\cite{HITACHI2005247}, and attempted to expand the work of Lindhard et~al. to atomic gases and molecular gases, and their binary mixtures~\cite{HITACHI20081311}.

In this article, ionisation quenching factors are estimated for a range of gaseous targets by exploiting available measurements of the mean ionisation energy $W$ for electrons and ion species. Measurements of $W$ have been the topic of intense investigations for over a century~\cite{icru31.1979,international1995iaea}, with improved experimental methods and increased precision over time.

\section{Theoretical background} 
\label{sec:theory}
In the following, a brief description of the theoretical
investigations on $W$ is provided and the connection to the ionisation quenching factor is discussed.

\subsection{Description of the W-value}
The average energy expended by ionising radiation to produce an electron-ion pair 
inside gaseous and liquid media, or an electron-hole pair in solids, is denoted by $W$. When all of the kinetic energy $E$ of the ionising particle is deposited in a medium the $W$-value is:
\begin{equation}
\label{W_def}
    W = E/N_i
\end{equation}
where $N_i$ is the mean number of electron-ion (or electron-hole) pairs created. It is assumed that any secondary particles produced during
the energy loss process, e.g. $\delta$-electrons,
bremsstrahlung photons, or electron and nuclear recoils, also dissipate
their energy in the medium and the produced electron-ion pairs
are included in $N_i$~\cite{icru31.1979}.

The W-value contains all the required information for converting the kinetic energy dissipated by ionising radiation in a medium to electric charge, and vice versa. The magnitude of the W-value is the outcome of the competition between ionising and non-ionising processes, such as excitation of atomic electrons and molecular degrees of freedom, dissociation processes (neutral fragmentation) in molecular gases, and production of neutral atomic recoils. As a result the W-value also depends on the interacting particle species and, in general, tends to a constant value for increasing energy.

\subsubsection{The W-value in pure gases}
An intuitive explanation of the W-value magnitude for the simplest case of fast electrons in pure gases was provided by Platzman~\cite{Platzman1961}, based on energy balance arguments. The kinetic energy is apportioned in three parts:
\begin{inparaenum}[a)]
\item production of electron-ion pairs; 
\item production of discrete excited states or neutral dissociation fragments; and
\item production of sub-excitation electrons, with kinetic energy
  below the lowest electronic excitation energy of the medium. 
\end{inparaenum}

This relation is written as:
\begin{equation}
\label{eq:w_ener_bal_1}
E = N_{i}\bar{E}_{i} + N_{ex}\bar{E}_{ex} + N_{i}\epsilon
\end{equation}
where $\bar{E}_i$ is the mean
energy transfer per electron-ion pair, $N_{ex}$ is the mean number of excited
states produced, $\bar{E}_{ex}$ is the mean energy transfer per excited
state, and $\epsilon$ is the mean kinetic energy carried by
sub-excitation electrons. Dividing by $N_i$ results in: 
\begin{equation}
\label{eq:w_ener_bal_2}
W = \bar{E}_{i} + \frac{N_{ex}}{N_{i}}\bar{E}_{ex} + \epsilon
\end{equation}
The W-value is more naturally expressed relative to the first ionisation threshold $I$ of the gas:
\begin{equation}
\label{eq:w_ener_bal_3}
\frac{W}{I} = \frac{\bar{E}_{i}}{I} + \frac{N_{ex}}{N_{i}}\frac{\bar{E}_{ex}}{I} + \frac{\epsilon}{I} 
\end{equation}
Since every term on the right-hand side of Eq.~\ref{eq:w_ener_bal_3}
depends on the energy of the ionising particle, 
the W-value also exhibits an energy dependence, which becomes weaker for $E\gg I$. Platzman uses helium as an example to demonstrate the completeness of 
the approach, where 
the magnitude of each term for high energies is:
\begin{equation}
\label{eq:w_ener_bal_4}
1.71 \approx 1.06 + 0.85 \cdot 0.40 + 0.31
\end{equation}
Each term on the right-hand side of Eq.~\ref{eq:w_ener_bal_4} has been derived from methods other than the absolute measurement of the W-value, which is shown on the left-hand side. 
The magnitude of these terms provides a 
sense of the apportionment of the ionising electron's kinetic
energy. The value of 1.06 for the ratio $\bar{E}_i/I$ is greater than
one because of the production of excited and multiply charged ions.
This effect is present in all noble gases and is further
enhanced in molecules thanks to the larger number of available
degrees of freedom. The term $\bar{E}_{ex}/I$ is in general less than
one but close to it because the energy levels of most excited states lie below or close to $I$. The ratio $N_{ex}/N_{i}$ depends strongly on the atomic and molecular dynamics, taking values of approximately 0.5, 1, and 2 for closed-shell atoms, molecules, and  free radicals, respectively~\cite{Platzman1961}. Finally, the ratio $\epsilon/I$ is smaller than unity because by definition sub-excitation electrons have energy lower than $I$, and this is the part dissipated as heat. The value of approximately 0.3 applies to most noble gases but is less than that for molecules because molecular ions retain more energy than atomic ions~\cite{icru31.1979}. Overall, high energy ionising particles have a $W/I$ ratio of 1.7--1.8 in noble gases and 2.1--2.5 in molecular gases.

\subsubsection{W-value dependence on particle species and energy}
\label{sec:en_dep}
The W-value depends on the 
ionising particle species and energy. 
For electrons with energy lower than a few keV, the cross section for electronic excitation increases with respect to that for ionisation as the electron energy decreases~\cite{hatano2010charged}. The electron W-value as a function of kinetic energy is, typically,
described as:
\begin{equation}
\label{eq:w_ener_depend}
W(E) = \frac{W_{a}}{1-U/E}
\end{equation}
where $W_a$ is the asymptotic W-value for $E\gg I$ and $U$ is a
constant close to the average energy of sub-excitation
electrons~\cite{inokuti1975}. This relation is in agreement with
measurements in a number of gases including nitrogen, methane, and propane, 
down to low energies~\cite{Waibel1983, Combecher1980,
  waibel_1978}, although some measurements suggest a stronger
dependence~\cite{Cole1969}. This is discussed in detail in 
Refs.~\cite{icru31.1979, international1995iaea}.

The energy dependence of the W-value for charged hadrons,  
such as protons, $\alpha$-particles, mesons, and heavy ions is more complicated as it involves a larger number of reactions compared to electrons. During the energy loss process, 
in addition to secondary electron generation, electron capture and loss cycles are taking place. 
Moreover, collisions of ions with atoms and molecules result in energy transfers that may lead to further ionisation when ion velocities become comparable with electron orbital velocities. 
When ions carry electrons, as expected for slow ions, collisions with atoms and molecules give a number of products, including energetic ions and molecular fragments, that may result in further ionisation. 
A theory description that reliably predicts the W-value is not yet available and cross sections for the individual process are under investigation~\cite{evans2003atomic}. 

The energy balance formulation of Platzman can be extended for ions by adding the contribution of neutral recoils created during the cascade~\cite{lindhard1963integral}: 
\begin{equation}
\label{eq:w_ener_bal_ions}
E = N_{i}\bar{E}_{i} + N_{ex}\bar{E}_{ex} + N_{i}\epsilon + N_{nr}\bar{E}_{nr}
\end{equation}
Dividing by $N_{i}$ gives:
\begin{equation}
\label{eq:w_ener_bal_ions_2}
W_{ion} = \bar{E}_{i} + \frac{N_{ex}}{N_{i}}\bar{E}_{ex} + \epsilon + \frac{N_{nr}}{N_i}\bar{E}_{nr}
\end{equation}
where $N_{nr}$ is the number of neutral recoils produced and
$E_{nr}$ is their average kinetic energy. Other terms keep their original
meaning as in Eq.~\ref{eq:w_ener_bal_1}. The $N_{i}\bar{E}_{i}$ and $N_{ex}\bar{E}_{ex}$ terms include ionisation and excitation, respectively, induced by
the secondary ions in addition to that induced by the secondary electrons.

\subsection{Relation of the W-value and the ionisation quenching factor}
The W-values for different ionising radiations encapsulates all necessary information to estimate differences in the amount of induced ionisation. The quenching factor parametrises these differences. In Section~\ref{sec:outline}, three definitions of the quenching factor were outlined. The first  quenching factor definition is experimentally accessible only in the case where a detector is able to measure the total amount of energy deposited by an ion in electronic energy losses, ionisation and excitations, as a fraction of its total kinetic energy. The two other definitions can be measured with a detector relying on ionisation. 

For a particle of energy $E$ depositing all of its energy in the detector, there will be $N_{i}=E/W$ primary electrons produced. The detector will record a signal $S$ proportional to $N_{i}$ where the constant of proportionality $G$ includes the response of the read-out electronics and any electron multiplication. The detector can be calibrated with electrons of known kinetic energy in order to relate the signal recorded with the corresponding energy deposited in the detector, $S\rightarrow E$. In the case of electrons, the corresponding signal for a given energy deposited will be $S_e = G \cdot N_i^e = G \cdot E/W_e(E)$. For ions of the same kinetic energy the corresponding signal will be $S_i = G \cdot N_i^i= G \cdot E/W_i$, which, through the calibration, would be interpreted as an electron equivalent energy $E_{ee} = N_i^i \cdot W_e(E)$. $E_{ee}$ is the energy visible to the detector, thus, from Eq.~\ref{eq:definitionBQuenchingFactor} the quenching factor can be written as: 
\begin{equation}
\label{eq:qf_exp_w}
q_f(E) = \frac{E_{ee}}{E} = \frac{ N^{i}_{i}\cdot W_e(E)}{E} =\frac{W_{e}(E)}{W_i(E)}
\end{equation}

When the kinetic energy is over a few keV, far away from the energy region where the W-value for electrons is energy dependent, the quenching factor can be derived from the ratio of the asymptotic W-value for electrons $W_a$ over the W-value for ions $W_{i}$ in the same medium. This is the case in the second definition, and is equivalent to Eq.~\ref{eq:qf_exp_w} when $W_e \rightarrow W_a$. 

In the third definition of the quenching factor, the energy dependence of 
$W_e$ 
is retained, and thus accounts for the effective quenching of electrons at energies below a few keV. 

\section{Quenching factor estimation} 
\label{sec:results}
\begin{figure*}[t]
\centering
\includegraphics[width=1\linewidth]{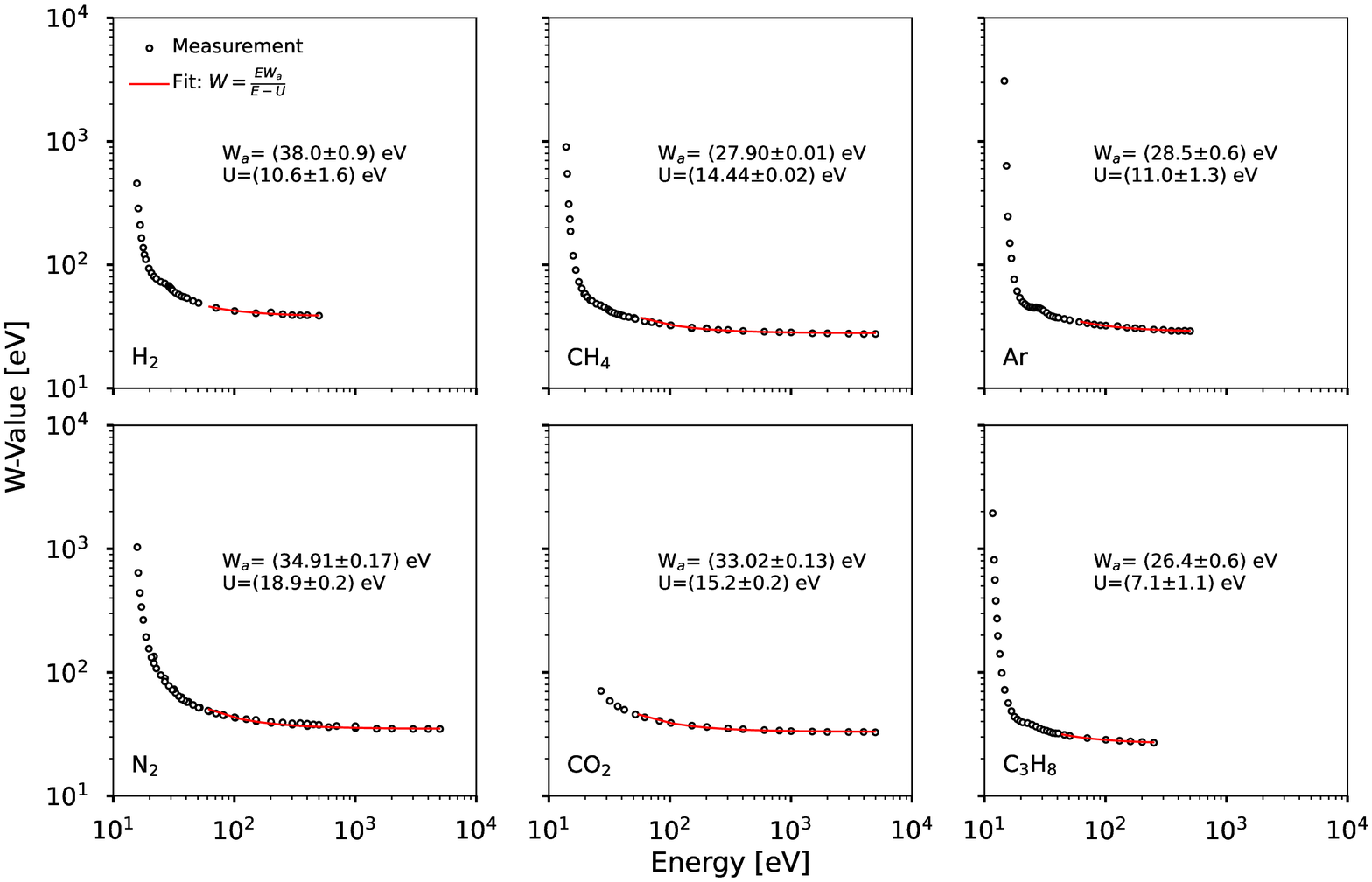}
\caption{W-values versus electron kinetic energy for the gases studied. Data from Ref.~\cite{Combecher1980} were used for H$_{2}$, Ar and C$_{3}$H$_{8}$. For CH$_4$, data from Refs~\cite{Combecher1980} and \cite{Waibel1983} are combined. The data from Refs~\cite{Combecher1980} and \cite{Waibel1983N2} were combined for $N_{2}$. Only the data from Ref.~\cite{Waibel1991CO2} was used for CO$_{2}$. The curves are fitted using Eq.~\ref{eq:w_ener_depend} to estimate the asymptotic W-value (W$_{a}$) for each gas.\label{fig:ElectronComparisonsLogLog}}
\end{figure*}

The estimation of the quenching factor, using Eq.~\ref{eq:qf_exp_w}, for different ions species relies on measurements of the W-values for ions and electrons, the latter used to obtain the asymptotic W-values for electrons. 
In this work, the measurements were selected based on two community reports produced by ICRU
(1971) and IAEA (1995) international commissions, which surveyed multiple studies of W-values.
The gases studied are H$_{2}$, CH$_{4}$, N$_{2}$, Ar, CO$_{2}$, C$_{3}$H$_{8}$ and W-values for electrons, protons, $\alpha$-particles, and
constituent ions are included. These gases 
are frequently used as counting gases and quenchers in proportional
counters~\cite{knoll2010radiation}, and as 
components of tissue-equivalent gases, used to emulate energy
deposition in the human body. Furthermore, these are of particular
interest as targets for rare event searches with gaseous
detectors.

\subsection{W-value measurements for electrons and ions in various gases}
The electron measurements by Combecher et~al.~\cite{Combecher1980}, Waibel
and Grosswelt~\cite{Waibel1983, Waibel1991CO2, Waibel1983N2} are presented in
Fig.~\ref{fig:ElectronComparisonsLogLog}, and 
demonstrate the discussed energy
dependence of the W-value. CO$_2$ data from Ref.~\cite{Combecher1980} were not used due to their discrepancy with other published measurements and calculations~\cite{international1995iaea}.
The data for each gas are fitted with Eq.~\ref{eq:w_ener_depend} using $W_a$ and $U$ as free
parameters to estimate the asymptotic W-value for electrons in each
gas. Only measurements for electron kinetic energies substantially larger than the first ionisation threshold are included in the fit. The measurement uncertainties are dominated by systematic effects, which are assumed to be fully correlated across the considered  energy range. These results are summarised in Table~\ref{tab:Wa-table}. 
The resulting asymptotic W-values appear 
higher than those
suggested by ICRU~\cite{icru31.1979} for high energy electrons.
The recommended ICRU values are an arithmetic average of
measurements published by a number of authors before 1971, using different energies and methodologies, and  the assigned uncertainties were chosen to embrace most of the available results. 
The results  presented here are based on the latest measurements recommended by the IAEA
report~\cite{international1995iaea}. The difference between the derived asymptotic W-value and the corresponding ICRU recommended value is assigned as a systematic uncertainty in the quenching factor estimate.
\begin{table}[]
\caption{W-values for electrons in various gases. The asymptotic W-values are derived from the fits shown in Fig.~\ref{fig:ElectronComparisonsLogLog}.\label{tab:Wa-table}}
\begin{tabular}{ccc}
  \hline
    Gas   & \multicolumn{2}{c}{W [eV]}        \\ 
          & ICRU           & Asymptotic      \\   \hline 
    H$_2$ & 36.5$\pm$0.7   &  38.0$\pm$0.9 \\
    CH$_4$& 27.3$\pm$0.6   & 27.90$\pm$0.01 \\
    N$_2$ &34.8$\pm$0.7    & 34.91$\pm$0.17\\
    Ar    & 26.4$\pm$0.5   & 28.5$\pm$0.6\\
    CO$_2$ & 33.0$\pm$0.7  & 33.02$\pm$0.13\\
    C$_3$H$_{8}$ & 24.0$\pm$0.5 & 26.4$\pm$0.6\\
 \hline
\end{tabular}
\end{table}

\begin{figure*}[t]
\centering
\includegraphics[width=1\linewidth]{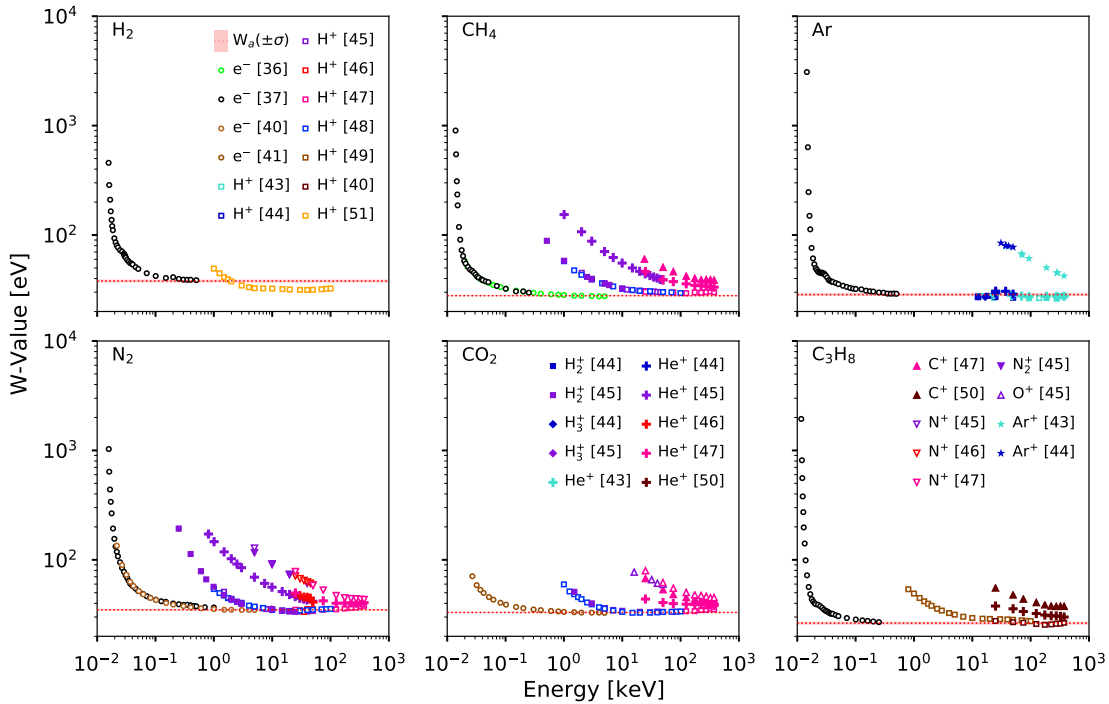}
\caption{W-value versus the kinetic energy of electrons and ions for gases used in this study. The asymptotic value estimated by the fit in Fig.~\ref{fig:ElectronComparisonsLogLog} is displayed by the dashed line with its statistical uncertainty given as a band.\label{fig:w-comp}}
\end{figure*}

The W-values for protons and heavier ions in the aforementioned
gases were taken from various sources. Specifically, Chemtob et~al.~\cite{Chemtob_1978} and
Phipps et~al.~\cite{phipps1964} were used for the W-values of protons,
H$_{2}^{+}$, H$_{3}^{+}$, He$^{+}$ and Ar$^{+}$ in Ar gas; Huber et~al.~\cite{huber_1985}, Boring et~al.~\cite{boring_1965}, Nguyen et~al.~\cite{Nguyen_1980}, Waibel et~al.~\cite{WaibelandWillems_1992} for
protons, H$_{3}^{+}$, He$^{+}$, N$^{+}$ and N$_{2}^{+}$ in
N$_{2}$ gas; Huber et~al.~\cite{huber_1985}, Nguyen et~al.~\cite{Nguyen_1980} and Waibel et~al.~\cite{WaibelandWillems_1992} for
protons, H$_{2}^{+}$, He$^{+}$, C$^{+}$ and O$^{+}$ in
CO$_{2}$ gas; Huber et~al.~\cite{huber_1985}, Nguyen et~al.~\cite{Nguyen_1980} and Waibel et~al.~\cite{WaibelandWillems_1992} for
protons, H$_{2}^{+}$, He$^{+}$, and C$^{+}$ in CH$_{4}$ gas; Willems et~al.~\cite{Willems_1998}, and Posny et~al.~\cite{Posny_1987} for
protons, He$^{+}$ and C$^{+}$ in C$_{3}$H$_{8}$ gas; Grosswelt et~al.~\cite{Grosswendt_1997} for protons in H$_{2}$ gas. The W-value for
electrons and ions in the gases studied are shown in
Fig.~\ref{fig:w-comp}. In the case of molecular ions the energy mentioned
is per atom. In the case of proton and H$_{2}$ curves, it is observed
that they, practically, coincide. This is explained by the molecules in
the ion beam dissociating in the initial collisions with the gas
constituents~\cite{huber_1985}.

\begin{figure}[!h]
\centering
\includegraphics[width=0.91\linewidth]{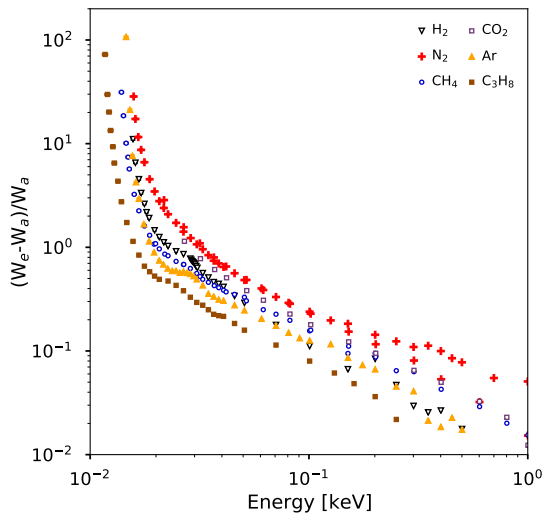}
\caption{Relative difference of measured electron W-values to the asymptotic W-value in the gases under study. 
\label{fig:WelectronDiffAsymptooticRatio}}
\end{figure}

In Fig.~\ref{fig:WelectronDiffAsymptooticRatio} the difference between the measured electron W-value to  the corresponding asymptotic value, normalised to the latter, is shown.

\subsection{Results}
\begin{figure*}[ht]
\centering
\includegraphics[width=1\linewidth]{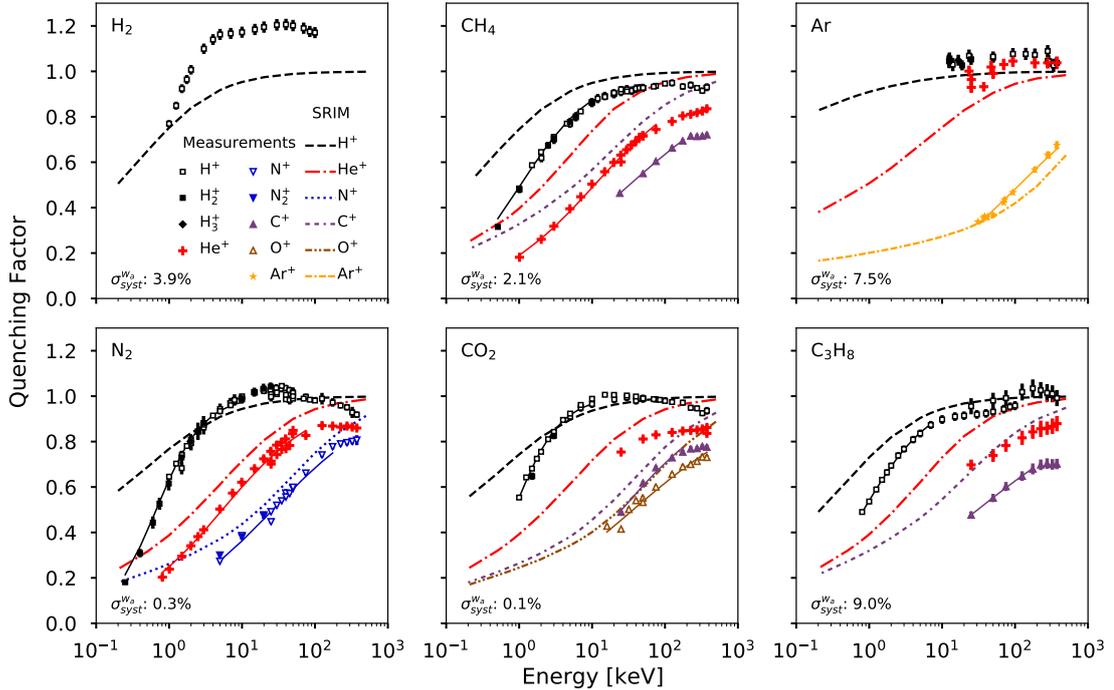}
\caption{Ionisation quenching factor versus ion energy for several
  gases. A fit to the data is overlaid (solid lines), as described in the text. The quenching factors estimated by SRIM are also provided for
  comparison. $\sigma^{W_{a}}_{syst}$ is the systematic uncertainty on W$_a$, which is taken to be the difference between the fitted and ICRU recommended W$_a$. \label{fig:qf-comp}}
\end{figure*}

The quenching factor for each ion species in a gas was estimated by dividing W$_{a}$ in that gas by the W-value of an ion species, as per Eq.~\ref{eq:qf_exp_w}. The results are displayed in Fig.~\ref{fig:qf-comp}. The uncertainties were calculated using the uncertainty in the estimation of the W$_{a}$ for electrons and the uncertainty for the W-values for ions provided in the respective publications. Where possible, the data were fit with a functional form inspired by Lindhard et~al.: \begin{equation}
q_{f}(E) = \frac{E^{\alpha}}{\beta + E^{\alpha}}\,,
\end{equation}
where $\alpha$, $\beta$ are free parameters. The data for different molecules of an atom are combined in the fit.
For reference, the estimated W-values for each ion species are presented along with the W-value provided by SRIM. 

\section{Discussion} 
\label{sec:discussion}
The quenching factor for gases, like the $W$-value, depends on the incoming particle species and kinetic energy. It is independent of the gas density over a wide range, up to the formation of electronic bands at elevated densities, for example at around $10\;\si{\bar}$ in the case of Xe~\cite{BOLOTNIKOV1997360, Aprile:2008bga}. 
This allows for $W$-values and quenching factors for gases to be estimated with specifically designed experiments, and then be applied to other experimental configurations in a wide range of densities. The method presented is based on this principle, utilising dedicated $W$-value  measurements to estimate the quenching factor in pure gases.

\subsection{Energy dependence of the quenching factor}
The results display the expected energy dependence of the quenching factor which is a consequence of the energy dependence of the $W$-value for ions. For electrons with high kinetic energy, typically above a few keV, the $W$-value is approximately constant, approaching the asymptotic value. However, for decreasing kinetic energy, the electron $W$-value exhibits a strong energy dependence, which introduces
an effective electron ionisation quenching. This additional reduction in ionisation has to be taken into account when converting between nuclear recoil energies and electron equivalent energies.
In order to accurately determine the electron equivalent energy of the ion, the energy dependent electron W-value should be taken into account. 
The difference between W$_{a}$ and the measured W-value for electrons in various gases is shown in Fig.~\ref{fig:WelectronDiffAsymptooticRatio}. At an energy of $500\;\si{\eV}$ there is approximately a $2-8\%$ difference between the two, which increases with decreasing energy. 

\subsection{Proton quenching factor greater than unity } 
\label{section:unity}
The W-value for high energy protons, $\mathcal{O}$(MeV) kinetic energy, in any gas is in general larger than or equal to the one for electrons in the same gas \cite{icru31.1979}. However, in  Fig.~\ref{fig:qf-comp}  the quenching factor for protons in H$_{2}$, Ar, N$_{2}$, and CO$_{2}$ is found to be greater than unity in some energies.
This counter intuitive effect is a result of the minimum in the proton W-value curve which is observed in those gases, and can be seen in Fig.~\ref{fig:w-comp}. This is attributed to charge exchange reactions involving H and the molecule (atom) of the gas~\cite{Nguyen_1980, Waibel_1992}. These reactions are dominant in the energy range around the minimum. Although commonly referred to as the ionisation quenching factor, a more appropriate term would be `ionisation yield', similar to the term `scintillation yield' used in scintillation detectors.

As an example the case of N$_{2}^{+}$ is discussed, while a detailed discussion can be found in Ref.~\cite{Edgar1973}.
The charge exchange runs in a cycle which starts with the proton capturing an electron resulting in the production of neutral H and a N$_{2}^{+}$,
\begin{equation}
H^{+} + N_{2} \rightarrow H + N_{2}^{+}
\end{equation}
The H atom can further undergo one of the following reactions:

\begin{eqnarray}
\label{eq:excit}
& H + N_{2} \rightarrow H + N_{2}^{*}\\
\label{eq:elast}
& H + N_{2} \rightarrow H + N_{2}\\
\label{eq:ioniz}
& H + N_{2} \rightarrow H + N_{2}^{+} + e^{-}
\end{eqnarray}
The H atom can go through the N$_{2}$ excitation
reaction in Eq.~\ref{eq:excit} or the elastic scattering
reaction in Eq.~\ref{eq:elast}, before ending the cycle by the ionisation
reaction in Eq.~\ref{eq:ioniz}. This 
cycle converts an initially pure proton beam into a mixture of H$^{+}$
and H, where the latter dominates for energies above approximately 10\,keV.
At low energies, the channels of N$_{2}$ ionisation and excitation
compete almost equally. The addition of secondary electrons, arising
primarily from the charge exchange reactions and the large number of ions
produced, results in an increased ionisation yield. Above about 100 keV
the charge exchange channel shuts off and results in a rise of the
W-value, which is lowered again near the MeV-scale 
due to an increase of
proton induced ionisation. The range at which this effect is dominant is observed to be gas dependent. 

\subsection{Comparison with predictions from SRIM}
In Fig.~\ref{fig:qf-comp} the estimated quenching factor curves are compared with quenching factor from SRIM calculations.  The level of agreement with SRIM varies and depends on ion type, energy range and gas. SRIM predictions seem to be in reasonable agreement with results for N$^{+}$ in N$_{2}$, C$^{+}$ and O$^{+}$ in CO$_{2}$ over 10 keV. SRIM predictions are also comparable to the results for He$^{+}$ in N$_{2}$, CO$_{2}$, Ar, as well as Ar$^{+}$ in Ar over 10 keV. Below 10 keV SRIM predictions are consistently higher than the estimated quenching factors. This effect may be attributed to the overestimation of the electronic stopping power in low energies by SRIM~\cite{paul2013}. 

\begin{figure}[!h]
\centering
\includegraphics[width=0.91\linewidth]{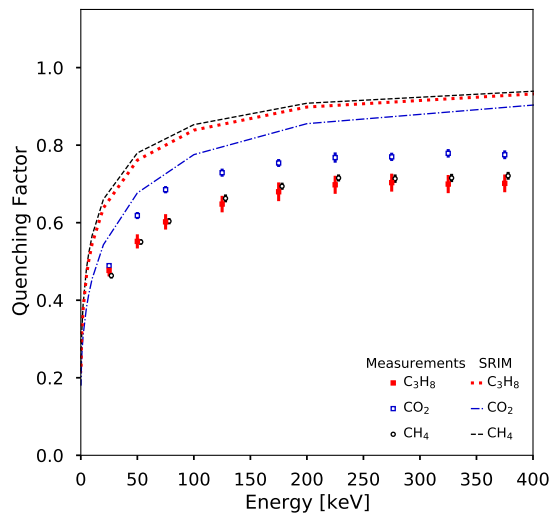}
\caption{Quenching factor for C$^{+}$ ions in CO$_{2}$, CH$_{4}$ and C$_{3}$H$_{8}$. The energy in CH$_4$ data has been increased by 3\;keV for visibility.\label{fig:CarbonComparison}}
\end{figure}

Two further points of disagreement between SRIM and the experimental results involve protons in all gases, except organic molecular gases, and all ion types in organic molecular gases. 
In the case of protons in H$_{2}$, Ar, N$_{2}$, and CO$_{2}$ the discrepancy  originates from the charge exchange cycle, discussed in section~\ref{section:unity}, resulting in additional ionisation with respect to the electronic stopping power estimated by SRIM. 
In hydrocarbons, as shown in Fig.~\ref{fig:CarbonComparison},
the order of quenching factors is inverted with respect to what would be anticipated by molecular mass arguments. For example, as predicted also by SRIM, ions in CO$_{2}$ would be expected to have a lower quenching factor than in CH$_{4}$ or C$_{3}$H$_{8}$. This behaviour has been previously reported~\cite{HITACHI20081311}, but its source remains unclear.
   
\subsection{Extension to other gases and gas mixtures}
The presented measurements were performed using precision experiments in the context of medical physics and dosimetry. As such, the gases studied in the literature are those with particular relevance to that field. However, the same measurement methods can be employed to study any gas of interest, particularly because the $W$-value measurements generally employ ionisation detectors. The method presented in this work can be applied to $W$-value measurements in other gases as they become available. Gases of particular importance are noble gases and molecular gases, such as CF$_4$, CH$_4$ and C$_4$H$_{10}$, as well as their mixtures. These are of interest to DM search experiments such as those mentioned in the introduction. Through dedicated measurements, the $W$-value could be extended to lower kinetic energies to better cover the region of interest for DM searches.
\section{Summary} 
\label{sec:summary}
Ionisation quenching is a crucial effect on the measurement of low energy ions in a number of applications, including direct DM searches. Due to the scarcity of measurements in relevant materials and energy ranges, one often has to rely on simulations, 
theoretical models, or extrapolation of measurements, leading to large uncertainties.

In this article, a distinction between the several definitions of the quenching factor found in literature is made, the 
importance of the increased electron W-value for low kinetic energies is highlighted, 
and estimates of the ionisation quenching factor for protons, $\alpha$-particles, and heavier ions are presented using measurements of their W-values and W-values of electrons in H$_{2}$, CH$_{4}$, Ar, N$_{2}$, CO$_{2}$, and C$_{3}$H$_{8}$ gases.
The obtained quenching factor estimates are discussed in terms of ion species and kinetic energy in a given gas, along with gas-specific effects, for example additional ionisation induced by charge exchange reaction cycles. The data are fit with a functional form inspired by Lindhard et~al., however, no extrapolation to lower energies has been attempted. In the future, dedicated measurements will provide better coverage of the region of interest for DM searches, thus reducing a major systematic uncertainty for these experiments.  Moreover, comparisons with simulations are included, where such effects are not currently included for the calculation of the electronic stopping power.

The presented method uses data from dedicated high-precision experiments and is directly applicable to any other gas for which W-value measurements become available. Since the quenching factor is an intrinsic property of a given gas, obtained estimates may be applied to any measurement using said gas. In the future, dedicated measurements can be performed using similar experimental set-ups to explore the lowest energy ranges and other pure gases and mixtures. These data could then be compared to phenomenological and theoretical models to provide a tool for estimating the quenching factor.

\section*{Acknowledgements} \label{sec:acknowledgements}
This work was performed in the context of the NEWS-G experiment. This project has received funding from the European Union's Horizon 2020 research and innovation programme under the Marie Sk\l{}odowska-Curie grant agreement no 841261 (DarkSphere).

\bibliography{mybiblio}

\end{document}